\documentclass[
	pra,
	aps,
	twocolumn,
	showpacs,
	groupedaddress,
	superscriptaddress]
{revtex4-1}
\usepackage{graphicx}
\usepackage{epstopdf}

\begin{document}

\newcommand{\be}{\begin{equation}}
\newcommand{\ee}{\end{equation}}
\newcommand{\beq}{\begin{eqnarray}}
\newcommand{\eeq}{\end{eqnarray}}

\newcommand{\w}{\omega}
\newcommand{\W}{\Omega}
\newcommand{\g}{\gamma}
\newcommand{\G}{\Gamma}
\newcommand{\E}{\hat{\cal E}}
\renewcommand{\a}{\hat a}
\renewcommand{\Im}{\mbox{Im}}
\newcommand{\s}{\sigma}
\newcommand{\n}{\bar{n}}
\newcommand{\ket}{\rangle}
\newcommand{\bra}{\langle}
\newcommand{\nn}{\nonumber}
\providecommand{\kb}[2]{\lvert\,#1\rangle\langle#2\rvert}
\newcommand{\me}{\mathrm{e}}
\newcommand{\dif}{\mathrm{d}}
\newcommand{\bnn}{\begin{eqnarray*}}
\newcommand{\enn}{\end{eqnarray*}}

\newcommand{\ds}{\displaystyle}
\newcommand{\dd}{\partial}
\newcommand{\dt}{\ds\frac{\dd}{\dd t}}
\newcommand{\dz}{\ds\frac{\dd}{\dd z}}
\newcommand{\D}{\ds\left(\frac{\dd}{\dd t} + c \frac{\dd}{\dd z}\right)}
\newcommand{\nt}[1]{{\it #1}}
\newcommand{\Rb}{Rb}
\newcommand{\Fig}{Fig.\ }

\newlength{\textwidthm}
\setlength{\textwidthm}{\columnwidth}
\addtolength{\textwidthm}{-\parindent}
\addtolength{\textwidthm}{-\parindent}

\newlength{\texthwidthremark}
\setlength{\texthwidthremark}{\columnwidth}

\newcommand{\remark}[1]{\\ \hspace*{-.5cm} 
\parbox[t]{\texthwidthremark}{
{\large\it #1\\}}
}




\title{Cooperative parametric resonance of the spin one half system of the dense atomic gas\\
}

\author{Yuri~V.~Rostovtsev}
\affiliation{
	Department of Physics,
	University of North Texas,
	Denton, Texas 76203, USA
}

\author{Vladimir~L.~Safonov} 
\affiliation{
Mag and Bio Dynamics, Inc., Granbury, Texas 76049, USA
}
\affiliation{
Physical Science Department, Tarrant County College - South Campus, Fort Worth, Texas 76119, USA
}

\author{Marlan~O.~Scully}
\affiliation{
Department of Physics, Texas A\&M University, College Station,  Texas 77843, USA
}
\affiliation{
Department of Mechanical and Aerospace Engineering, Princeton University, Princeton, New Jersey 08544, USA
}
\affiliation{
Department of Physics, Baylor University, Waco, Texas 76706, USA
}

\date{\today}

\begin{abstract}

The cooperative resonance for a spin one half system interacting with 
dc and ac magnetic field is considered. This interaction in the system collective 
regime can result in parametic resonance and rapid excitation of the excited spin state
of the dense atomic gas.  The phenomenon is studied using the density matrix approach.
We discuss the implementation of this effect and possible applications of the quantum amplification 
by superradiant emission of radiation. 

\end{abstract}


\maketitle

\newpage

\section{Introduction}

The paradigm of two-level systems in an electromagnetic field plays an extremely
important role in two quite different branches of physics, quantum optics and magnetic resonance theory. 
Using mathematical analogy, one can transfer the idea of the effect developed
in one branch to another branch and a similar effect to predict. 
Thus, we have an exciting opportunity and playground for the implementation of 
proof-of-principle model experiment to demonstrate new effect in the branch
where experimenting is easy. 
In this paper we analyze cooperative resonance in a spin one half system 
interacting with the magnetic field and demonstrate that this system can
exhibit superradiant emission of radiation similar to that of recently discovered 
for the two-level atomic gas. 

The so-called, quantum amplification by superradiant emission of radiation, 
 a new concept of coherent radiation generation, 
was proposed in Refs.\cite{svidzinzki2013,genkin00pra,mos}. 
It is based on the cooperative effects in an ensemble of two-level atoms
interacting with a radiation field in the resonator cavity~\cite{book0,book}.
In order to explain the idea of new effect, we write the density matrix
equations as
\beq
\label{eq1}
\dot\rho_{ab} = i(1 - 2\rho_{aa})\W_s,\\
\label{eq2}
\dot\W_s = i\W_a^2\rho_{ab}. 
\eeq
Here $\W_s =\wp E_s/\hbar$ is the Rabi frequency of the radiation field coupled to the
atomic ensemble, $\rho_{aa}$ is the atomic population in the excited state, 
$\rho_{ab}$ is the atomic coherence, and  $\W_a$ is the cooperative frequency
defined by
$$
\W_a^2=\frac{2\pi}{\hbar}\w_{ab}\wp^2 N,
\label{Wa}
$$ 
$N$ is the density of atoms. 

We can represent Eqs. (1) and (2) in the following form:
\be
\label{eq3}
\ddot\W_s + \W_a^2(1 - 2\rho_{aa})\W_s = 0.
\ee
When two-levell atoms are in the excited state ($\rho_{aa}>1/2$), 
their transition to the ground state leads to the generation of 
a superradiant pulse~\cite{dicke}. 

If the atoms are close to the ground state ($\rho_{aa}<1/2$), then 
 Eq.(\ref{eq3}) describes a harmonic oscillator.
So far as the energy
$$
{\cal E} = \left(\hbar\w_{ab}N\rho_{aa} + {|E_s|^2\over 4\pi}\right)V_{cavity}
$$  
 stored in the atoms and the laser field is conserved, 
the radiation oscillates at the cooperative frequency $\W_a$ 
and the energy  goes from the radiation to excite atoms and back. 

The amplitude of this oscillation rapidly increases
if the atomic population is modulated as 
$\rho_{aa} = \rho_{aa}^0 + \delta\rho_{aa}\cos(2\nu t)$
near the cooperative frequency $\nu \approx \W_a$. 
In this case the equation for radiation (\ref{eq3}) becomes the Mathieu equation, 
and we have 
the parametric resonance leading to the growth of the oscillation field amplitude.
The energy increases due to interaction of the atomic gas 
with the external modulation field, which leads 
to the population excitation and simultaneously increases the laser field. 

The above simple consideration showed that the cooperative resonance
 is a promising tool for the development of new sources 
for the coherent radiation generation.
However, in order to build more realistic approach 
in the optical range, we have to take into account the velocity distribution of atoms.
The cooperative frequency is not well-defined in this case
and to observe the cooperative effects, some additional conditions are required.
At the same time, we can develop similar models in the RF (or microwave) range. 
The interaction of intense ultrashort pulses with atomic system can
be studied with RF pulses as a model system~\cite{HLi10prl}. 
We acknowledge also the ground-breaking experiments
performed with RF radiation  \cite{cohentannoudji}. 
Such experiments in the RF region might furnish physical insight for the
development of approaches using the cooperative resonances. 

In this paper, we consider the cooperative resonance in the spin one half system 
coupled to the magnetic ac and dc field. The resonance with the cooperative 
frequency creates possibility to generate the magnetic field as well as 
to excite the system of spins. 
In particular, we consider a gas of two-level atoms 
in the presence of coherent driving field. The coherent field 
either has modulation at the cooperative frequency 
or detune off the resonance at the amount of the cooperative 
frequency. 


\begin{figure}[tb] 
\center{ 
\includegraphics[width=0.45\textwidth]{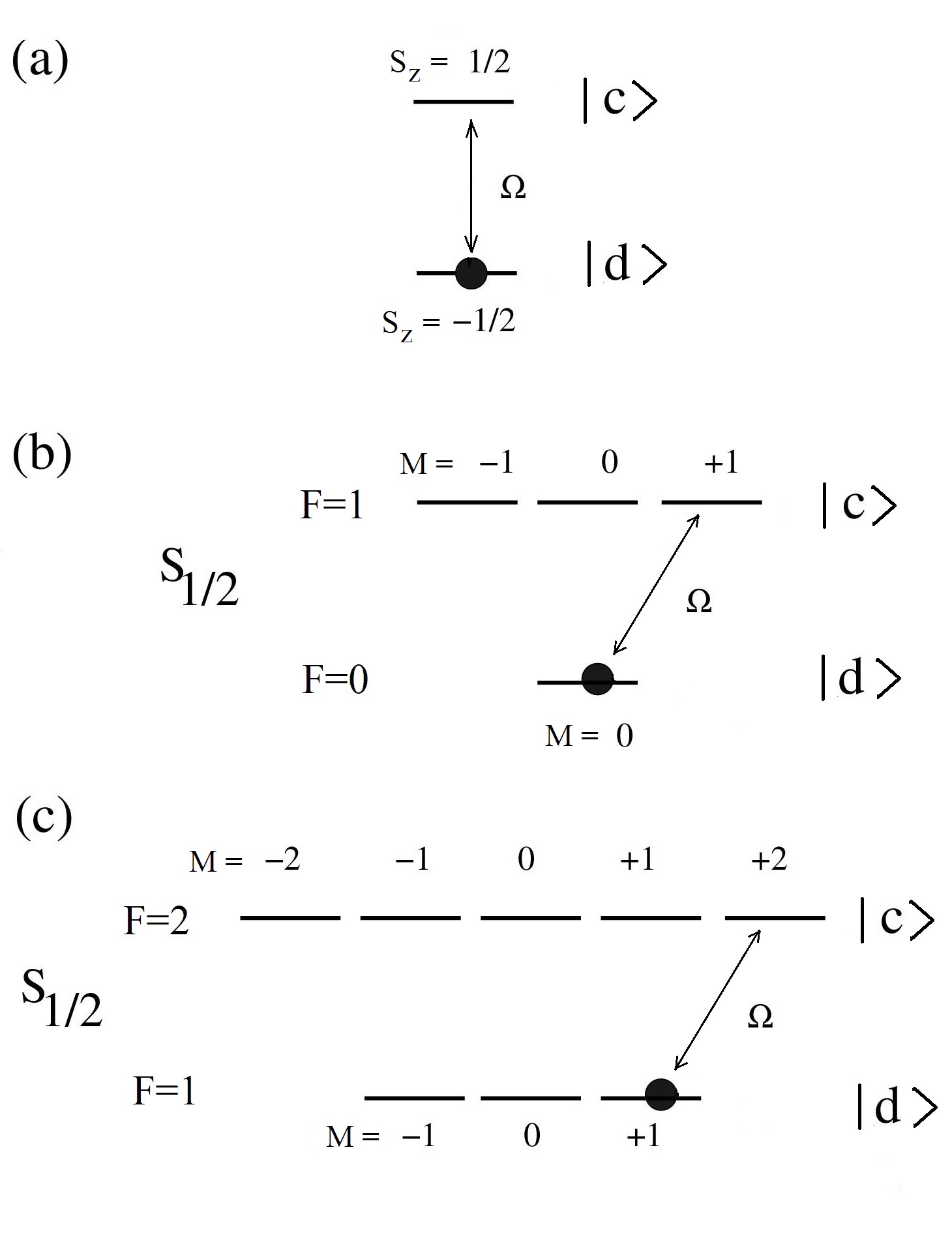}        
}
\caption{Level structure of (a) spin-1/2 system, (b) hydrogen atom, and (c) Rb atom. 
The population is optically pumped to the state $|d\ket$, namely:
(a) $|-1/2\ket$,  (b) $|F=0, M=0\ket$, and (c) $|F=1, M=+1\ket$.
}
\label{rb-level}
\end{figure}

\section{Model}

In this section we consider a two-level system, spin-1/2,  interacting with the circularly 
polarized magnetic field. Schematically this simple system is shown in
 Fig.~\ref{rb-level}a, and the basic equations are given in Appendix A.
 
In addition, we consider two examples of real two-level system in 
application to atoms, namely, hydrogen atom (see, Fig.~\ref{rb-level}b) and
 Rb atom  (see, Fig.~\ref{rb-level}c). 
 
One can optically pump all population to the particular
state $|d\ket$ (marked in Fig.~\ref{rb-level}a-c by a bullet)
and then apply the circular polarized RF field
between two levels as shown in the same figure. 
The frequency of the RF transitions 
can be controlled by a dc magnetic field. 

The experiment can be performed with RF (or, microwave) fields.
Note that another possible realization of the present results involves
experiments with atoms in Rydberg states, as in \cite{kleppner},
or using a RF field resulting in magnetic dipole transitions between levels 
with the same $F$ and different $M$ in the magnetic field 
\cite{cohentannoudji}. 

The spin Hamiltonian in the circular magnetic field $B_x= B_s\cos(\nu t)$ and $B_y
=-B_s\sin(\nu t)$, is given by
\be
{\cal H} = \mu_x B_x + \mu_y B_y = 
{B_s\over2}[\mu_+ e^{-i\nu t} + \mu_-e^{i\nu t}],
\ee
where 
$\mu_\pm = (\mu_x \pm i \mu_y)/\sqrt{2}$ and
${\vec{\mu}} = \mu_B({\bf L} + 2{\bf S})$ is the magnetic moment. 

Evolution of the state vector 
\be|\Psi\ket = \sum_{FM}a_{FM}e^{-i\w_F t}|FM\ket,\ee 
is determined by the equation  
\begin{widetext}
\beq
i\dot a_{FM} = 
{B_0\over2\hbar}\sum_{F'M'}
\left(\bra FM|\mu_+|F'M'\ket \exp[-i(\nu+\w_{F'F} t)] + 
\bra FM|\mu_-|F'M'\ket \exp[i(\nu-\w_{F'F} t)]
\right) a_{F'M'}
\label{eq-rb}
\eeq
\end{widetext}
The elements of matrices $\bra FM|\mu_\pm|F'M'\ket$
and the explicit form of Eq.~(\ref{eq-rb}) 
can be found in Appendix B.
To find the magnitization of the spins, we 
solve the set of Eqs.~(\ref{eq-rb}). 

We can also use the following equations of motion for
the density matrix $\rho$ at the times much shorter than the relaxation 
times:
\beq
\label{7}
\dot\rho_{aa} = 2\W_s \Im{\rho_{ab}}, \\ \nonumber
\dot\rho_{ab} = i(1 - 2\rho_{aa})\W_s.
\eeq

\begin{figure}[ht]
\begin{center}
\includegraphics[angle=0, width=0.75\columnwidth]
{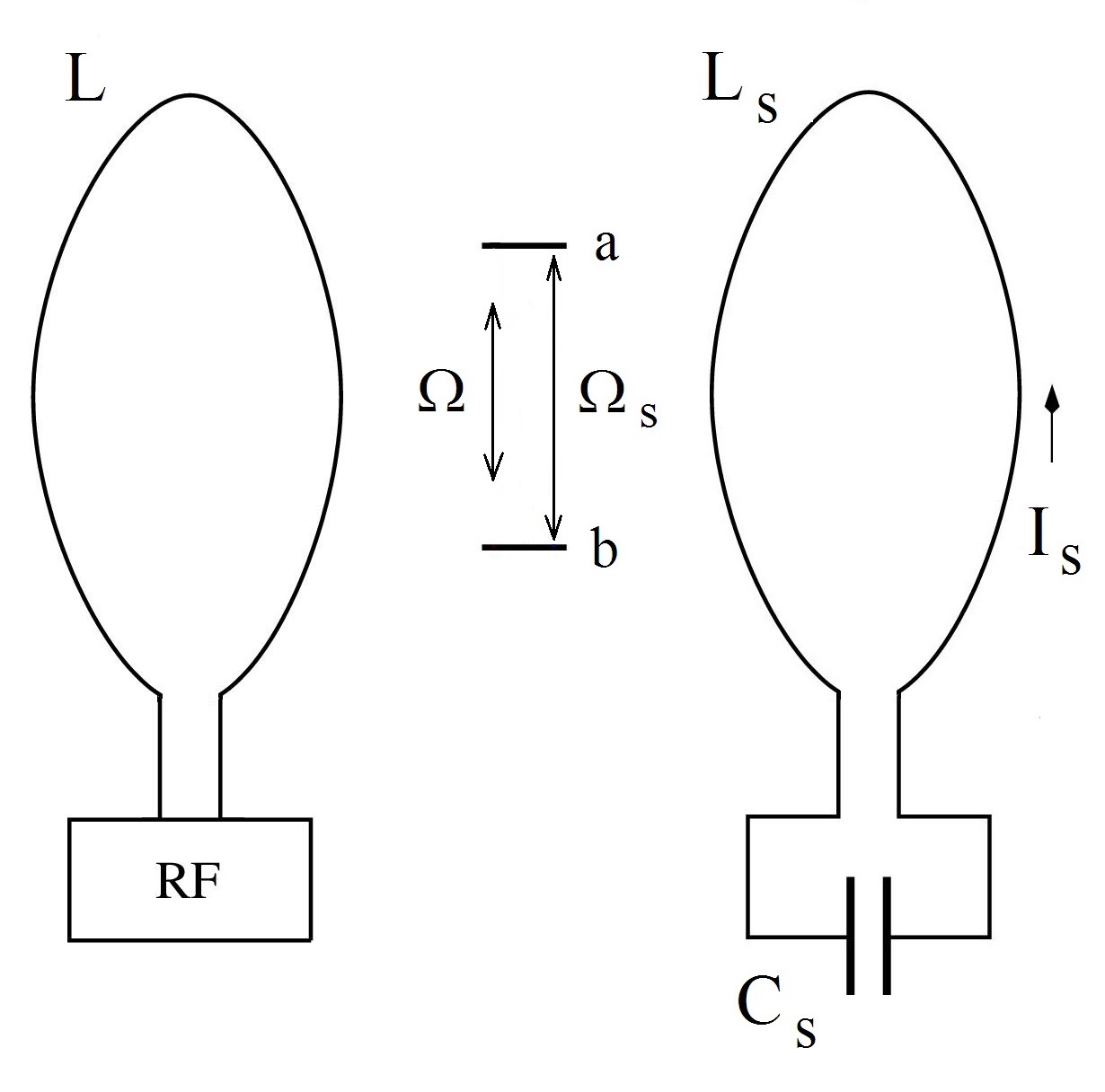}
\end{center}
 \caption{
The configuration of the Zeeman sublevels driven by RF field $\W$
created by RF generator, 
and the RF field $\W_s$ generated in a resonant contour.
\label{RF}
 }
\end{figure}

Now we turn our attention to the coupling of the induced magnetic polarization to 
the probe magnetic field $B_s$ that is created by 
the resonant RF circuit (with high quality factor Q), which consists of 
the inductance $L_s$ and a capacitor $C_s$ (see, Fig.~\ref{RF}) 
and has a resonant frequency at the frequency of the 
atomic a-b transition, is to be excited by Rb atoms.

It is possible to  consider a proof-of-principle experiment to demonstrate 
the mechanism of radiation generation. Using the Zeeman splitting of 
hyperfine magnetic sublevels, one can drive the system with a detuned RF field 
(see Fig.~\ref{RF}, the RF generator drives current $I$ through the coil 
to create a magnetic field $B$. This field is defined by the relation
 $L I=A_d B$, where $L$ is the inductance and $A_d$ is the 
area of the coil.


The oscillating at the frequency $\w$ 
magnetic dipole creates the electric field that
in the case of $kr\ll 1$ is given by{\cite{Jackson}}
\be
\vec E = -{\mu_0k^2\over 4\pi}\vec n\times\vec\mu 
{e^{ikr}\over r}\left(1+{i\over kr}\right)
\simeq -i{\mu_0k\over 4\pi}{\vec n\times\vec\mu\over r^2}, 
\ee
where $\mu_0$ is the permeability of vacuum, $k=\w/c$,  
$\vec\mu$ is the oscillating magnetic moment, 
\be
\vec\mu=\vec\mu_{ab}{\cal N}\rho_{ab}e^{-i\w t},
\ee 
${\cal N} = NV_{\mbox{sample}}$ is the number of the atomic spins, which
depends on the spin density $N$ and the volume of the sample. 
For simplicity, let us consider the magnetic spins being in the center of 
a circle wire, and, then, the electromotive force generated in 
the electric circuit is given by
\beq
V =\oint_{{\cal L}_s} \vec E\cdot\vec{dl}= 
-i{\mu_0k\over 4\pi}\oint_{{\cal L}_s}
{(\vec n\times\vec\mu)\cdot\vec{dl}\over r^2} \\ \nonumber
=-i{\mu_0k\over 4\pi}\vec\mu\cdot\oint_{{\cal L}_s}
{\vec{dl}\times\vec n\over r^2}=
-i{\mu_0k\over 4\pi}\vec\mu\cdot\vec{\cal J},
\eeq
where
\be
\vec{\cal J} = \oint_{{\cal L}_s} {\vec{dl}\times\vec n\over r^2}.
\ee
The coil current $I_s$ is then given by
\be
L_s\dot I_s + {q_s\over C_s} = V
\label{I_s}
\ee
and
\be
\ddot I_s + \w_s^2I_s = {\dot V\over L_s}, 
\label{ddI_s}
\ee
where $\w_s =({ L_sC_s})^{-1/2}$.

The current $I_s$ and the magnetic field $B_s$ created by the coil 
follow from the Biot-Savart law 
\be
\vec B_s = {\mu_0\over 4\pi}I_s\oint_{{\cal L}_s}\ds{\vec{dl}\times\vec r
\over r^3}={\mu_0\over 4\pi}I_s\vec{\cal J}.
\ee
One can write 
\be
\ddot {\vec B_s} + \w_s^2 \vec B_s = 
{\mu_0\over 4\pi}{\dot V\over L_s}\vec{\cal J}=
\ds{\dot V\over 2a_sL_s},
\ee
or,
\be
\ddot {\vec B_s} + \w_s^2 \vec B_s = 
-i{k\over L_s}\left({\mu_0\over 4\pi}\right)^2
(\dot{\vec\mu}\cdot\vec{\cal J})\vec{\cal J}.
\ee

Using the slowly varying amplitude approaximation, 
$B_s = \tilde B_se^{-i\w_s t}$, we obtain
\be
-2i\w_s\dot{\tilde B}_s = 
-i{k_s\over L_s}\left({\mu_0\over 4\pi}\right)^2
(\dot{\vec\mu}\cdot\vec{\cal J})\vec{\cal J}
\ee
where $k_s = \w_s/c$ and 
\be
\dot{\tilde B}_s = -i{\w_s\over 2cL_s}\left({\mu_0\over 4\pi}\right)^2
(\vec\mu_{ab}\cdot\vec{\cal J})\vec{\cal J}{\cal N}\rho_{ab}.
\ee
Introducing $\W_s = {\vec\mu_{ab}\cdot\tilde B_s/\hbar}$, one has
\be
\dot\W_s = -i{\w_s\over 2\hbar cL_s}
\left({\mu_0\vec\mu_{ab}\cdot\vec{\cal J}\over 4\pi}\right)^2
{\cal N}\rho_{ab} = -i\W_a^2\rho_{ab},
\label{Ws}
\ee
where
\be
\W_a^2 = {\w_s\over 2\hbar cL_s}
\left({\mu_0\vec\mu_{ab}\cdot\vec{\cal J}\over 4\pi}\right)^2
{\cal N} = 
{\w_s\mu_0^2\mu_{ab}^2{\cal N}\over 8\hbar cL_sa_s^2},
\ee
and $a_s$ is the radius of the coil.


\section{Cooperative parametric resonance}
\label{CR}

Let us consider theoretically the interaction of strong coherent field 
effects on the population
of two atomic spin states of a dense atomic gas interacting in a collective 
regime. It is an interesting way of generation of a laser field 
that is not based
on population inversion, but rather on the cooperative interaction 
with the ensemble of two-level atoms. 
The key feature of the approach is 
that the laser field is generated together with the population 
in the excited state.


To demonstrate the cooperative parametric resonance, 
we write the Rabi driving frequency in the rotating wave approximation as 
\be
\W = \W_0\cos(\nu t). 
\label{W}
\ee 
Introducing the phase
\be
\theta = \int^t_0 [ \W_s + \W_0\cos(\nu t) ]dt,
\label{theta}
\ee 
one can represent solutions of  Eqs. (\ref{7}) in the following form:
\beq
\rho_{aa} &=& \sin^2(\theta), \\
\rho_{ab} &=& {i\over 2}\sin(2\theta). 
\eeq
For a weak field one has  $\theta\ll 1$ and therefore,  $\rho_{aa} \simeq \theta^2$. 
In the beginning of lasing, we can neglect the change in phase due to
the laser field: $\W_s t \simeq 0$. 
Then the population of the excited state can be expressed as
\beq
\rho_{aa} \approx  {\epsilon^2\over2}[1- \cos(2\nu t)],
\eeq
where $\epsilon = \W_0/\nu$.

For the laser field Eq.(\ref{Ws}), we obtain 
\be
\ddot \W_s = - \w_0^2(1 + \epsilon^2\cos 2\nu t)(\W_s + \W_0\cos\nu t)
\label{Mathiew}
\ee
where $\w_0^2 = \W_a^2(1-\epsilon^2)$.

Let us try solution of Eq.(\ref{Mathiew}) in the form  
\be
\W_s = A_1 e^{i\nu t} + A_2 e^{-i\nu t}.
\ee
Here $A_{1,2}$ are slowly varying amplitudes for which we get 
\beq
\dot A_1 =  i\Delta A_1 - i G A_2 - i{\w_0^2\W_0\over2\nu},\\
\dot A_2 = -i\Delta A_2 + i G A_1 + i{\w_0^2\W_0\over2\nu},
\eeq
where $\Delta=(\nu^2 - \w_0^2)/2\nu$  and $G = \epsilon\w_0^2/2\nu$. 

Now we are looking for the solution of the form 
 $A_{1,2} \propto \exp(\lambda t)$ and obtain the following 
characteristic equation 
\be
(\lambda -i\Delta)(\lambda +i\Delta) - G^2 = 0.
\ee
The obtained characteristic numbers are:
\be
 \lambda = \pm\sqrt{G^2-\Delta^2}. 
\ee

Thus, for $\Delta = 0$, one can write the solution as
\beq
\left( \begin{array}{c} A_1 \\ A_2 \end{array} \right) =
{i-1\over 2G\nu}\w_0^2\W_0
\left( \begin{array}{c} 1 \\ i \end{array} \right)
(1 - e^{-G t})\\ +
{i+1\over 2G\nu}\w_0^2\W_0
\left( \begin{array}{c} 1 \\ -i \end{array} \right)
(e^{G t} - 1),  \nonumber
\eeq
that exhibits an exponential growth due to 
parametric resonance~\cite{Landau89}. 

\begin{figure}[ht]
\begin{center}
\includegraphics[angle=0, width=0.8\columnwidth]
{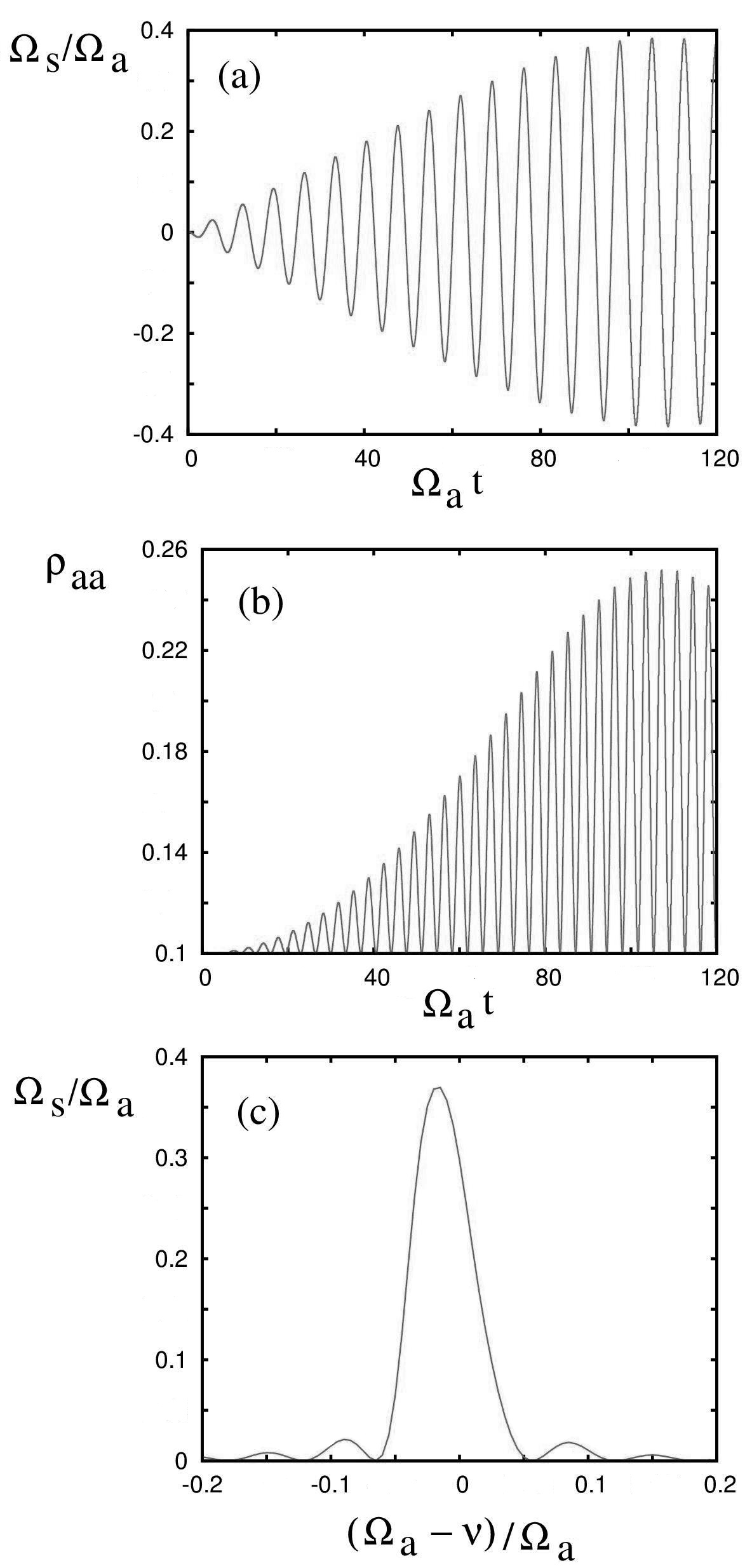}
\end{center}
 \caption{
(a) The growth of the probe field $\W_s/\W_a$ versus dimensionless time  $\W_a t$.
(b) The growth of the population in the excited state $\rho_{aa}$
 versus dimensionless time  $\W_a t$.
(c) The dependence of the growth of the probe field $\W_s/\W_a$ 
vs the detuning $(\W_a - \nu)/\W_a$.
 For simulation, the initial population 
in the excited state is $\rho_{aa}(0)=0.1$, and 
the drive amplitude is $\W_0  = 0.01\W_a$. 
\label{results}
 }
\end{figure}

To support the simplified calculations, we performed simulations of 
the set of Eq.(\ref{7}) and Eq.(\ref{Ws}). The results are shown 
in Fig.~\ref{results}.

The parametric resonance leads to the growth of the probe 
field $\W_s/\W_a$ in the case of modulation of the atomic population 
in the excited state is shown vs $\W_a t$.

The excitation occurs at the cooperative 
resonance ($\nu \simeq \W_a$). As can be seen from Fig.~\ref{results}, 
using a weak driving field $\W_0=0.01\W_a$, 
it is possible to excite the spin magnetization at almost the same 
the level ($\W_a\simeq 0.4\;\W_a$) as it would occured with superradiant 
generation corresponding to practically total population inversion.

This results can be observed in the experimental setup similar 
to the one described in~\cite{HLi10prl}. 
Using a longitudinal magnetic field $B=500$ G, the splitting is 
$\w_{ab} \simeq 318.3$~MHz. Then, for the atomic density $N = 10^{13}$ cm$^{-3}$, 
the cooperative frequency is $\W_a = 10^5$~s$^{-1}$, which 
is much larger than the ground state relaxation $\g_0 = 3\cdot 10^3$~s$^{-1}$
($\W_a \gg\g_0$).

The obtained above results can probably be applied to the nuclear spin
systems (see, e.g., Ref.~\cite{NMR}). 
It provides a new approach to the detection of nuclear  magnetic resonance. 
Usually, NMR is detected by the 
measuring of additional losses in the coils at the nuclear. 
Meanwhile, the current approach allows one to induce the magnetic polarization 
the population difference is given by 
$\Delta N_{ab} = N_0{\Delta E/ k_B T}$ where 
the density is $N_0 \simeq 10^{23}$~cm$^{-3}$
the cooperative frequency is $\W_a = 10^4$~s$^{-1}$.  
Then, modulation at the frequency $\Delta = \W_a$ leads to 
the strong magnetization of the nuclear transition.

We should mention another interesting opprotunity to find 
analogies (and therefore enrich our approach) using magnetic dynamics theory.
The point is that the theory of parametric resonance of magnons, quanta of spin waves
in magnetoordered systems is well-developed ~\cite{safonov}.
So far as the roles of nonlinear medium and resonator cavity are theoretically well clarified 
in this branch of physics, a similar structure of dymamic equations gives useful hints
how to update our simple model introducing new important parameters.

\section{Discussion}

In the paper, we study the new way of generation of coherent field based 
on the cooperative resonance in the system of spins which we consider as
a good way to implement a proof-of-principle experimental realization 
as well as probably new way of nuclear magnetic resonance detection.

This new approach is not based on the population inversion which 
is required for lasing, as it is well-known to implement lasing, 
population inversion usually is needed to overcome 
stimulated absorption~\cite{book0}. Also the technique is not related to 
the concept of lasing without population inversion (LWI)~\cite{book} that 
appears as a result of coherent 
effects~\cite{ok86jetp,fleisch05rmp,sau05pra,CPT,book,harris97phys2day} 
in atomic or molecular media~\cite{lwi1,lwi2,lwi3}. 
The LWI was demonstrated experimentally~\cite{lwi1e,lwi2e}, and 
it was even shown that lasing can exist without any inversion in any reservoir, 
even under thermodynamical equilibrium~\cite{ok98pra}. 

The physics of new way of generation is closely related to the cooperative 
response of quantum ensemble that are all practically in the ground state. 
Because of collective motion of the spin excitation 
at the cooperative frequency, the system undergoes the process of coherent 
excitation together with generation of coherent field. The physics is closely 
related to the so-called Dicke supperradiance~\cite{dicke} which is usually 
related to the spontaneous emission in the excited medium. 
Usually, spontaneous emission is an incoherent process that leads to 
relaxation of excitation in media. But in sufficiently dense media,
the spontaneous emission becomes a collective process as was shown by 
Dicke~\cite{dicke}. The cooperative behavior of $N$ atoms 
can speed up the relaxation processes, leading to a burst of radiation 
with an intensity proportional to the square of the number of atoms $N^2$. 
Recently, in a single photon superradiance~\cite{mos1, mos2, manassah08phl}, 
the collective Lamb shift was predicted and observed~\cite{mos3, LSh}.

In conclusion, we have studied and demonstrated several cases 
when the resonance with the cooperative frequency creates 
the possibility to generate coherent radiation. 
In particular, we consider a gas of two-level atoms in the presence of 
 either a CW or a pulsed coherent field that leads to substantial 
enhancement of the generated radiation under the condition that uses 
the cooperative frequency. 
A number of important applications of the generation of coherent radiation 
based on cooperative resonance phenomena that may lead to a significant 
progress in fundamental and applied physics.

\section*{Acknowledgments}

We thank Gombojav O. Ariunbold, Nikolai Kalugin, Jos Odeurs, 
Pavel Polynkin, David Lee, John Reintjes, Wolfgang P. Schleich, 
and Vlad Yakovlev for fruitful discussions, 
comments and suggestions, and gratefully acknowledge the support 
of the National Science Foundation Grants No. PHY-1241032
(INSPIRE CREATIV) and No. EEC-0540832 (MIRTHE ERC) 
and No. PHY-1068554, the Office of Naval Research,
and the R.A. Welch Foundation (Awards No. A-1261 and
No. A-1547).


%

\appendix

\section{Spin one half system} 

 The Hamiltonian for a spin-1/2 system in a magnetic field is defined by 
\be
\cal{H} = -{\vec\mu}\cdot  {\bf B},
\ee
where 
\be
{\vec\mu} = \mu_B (\sigma_x \hat{\bf x} +\sigma_y \hat{\bf y} +\s_z \hat{\bf z}).
\ee
Here the basis of two states, 
spin-up $|+\ket$  and spin-down $|-\ket$, is described by Pauli matrices 
\be
\sigma_x = 
\left( \begin{array}{cc}
0 & 1\\
1 & 0 
\end{array}
\right), \;
\sigma_y = 
\left( \begin{array}{cc}
0 & -i\\
i & 0 
\end{array}
\right), \;
\sigma_z = 
\left( \begin{array}{cc}
1 & 0\\
0 & -1 
\end{array}
\right).
\ee
The magnetic field has a dc longitudinal component $B_z$ and
the time-dependent transverse components 
\be
{\bf B} = B_z \hat{\bf z} +  B_s[\cos(\nu t) \hat{\bf x} + \sin(\nu t) \hat{\bf y}]. 
\ee
The longitudinal magnetic field causes splitting of the spin-up and spind-sown
states as
\be
{\cal H}_0 = \mu_B B_z \s_z = \mu_0 B_z\left( \begin{array}{cc}
1 & 0\\
0 & -1 
\end{array}
\right). 
\ee
The frequency, that corresponds to the transition between levels is defined by
\be
\omega_c = 2 \mu_B B_z /\hbar.
\ee
The transverse magnetic field causes coupling between these states
\be
{\cal V} = \mu_B B_s e^{i{H_0 t\over\hbar}}
[\s_x\cos(\nu t)) + \s_y\sin(\nu t)]e^{-i{H_0 t\over\hbar}}. 
\ee
Interaction Hamiltonian is given by 
\be
{\cal V} = -\mu_B B_s e^{i{H_0 t\over\hbar}}
\left( \begin{array}{cc}
0 & e^{-i\nu t}\\
e^{i\nu t} & 0 
\end{array}
\right)
e^{-i{H_0 t\over\hbar}} 
\ee
We see that the matrix elements, 
\beq
\bra +|{\cal V}|-\ket &=& -\mu_B B_s e^{-i(\nu-\w_c) t}, \\
\bra -|{\cal V}|+\ket &=& -\mu_B B_s e^{i(\nu-\w_c) t},
\label{lcp}
\eeq
coinside  with rotating wave approximation for the left-circularly polarized magnetic field. 
For the right-circularly polarized magnetic field the matrix elements are given by
\beq
\bra +| {\cal V}_{rcp}|-\ket &=& -\mu_B B_s e^{-i(\nu+\w_c) t}, \\
\bra -|{\cal V}_{rcp}|+\ket &=& -\mu_B B_s e^{i(\nu+\w_c) t}.
\label{rcp}
\eeq
Note that either the right-circularly or 
left-circularly polarized magnetic field is able to couple these two states. 

For the linearly polarized transverse magnetic field 
\be
{\bf B} = B_s{\bf x}\cos(\nu t)), 
\ee
the matrix elements have the following form: 
\beq
\bra +| {\cal V}|-\ket = -\mu_B B_s [e^{-i(\nu-\w_c) t} + e^{i(\nu+\w_c) t}], 
\label{counter1} \\
\bra -|{\cal V}|+\ket = -\mu_B B_s [e^{i(\nu-\w_c) t} + e^{-i(\nu+\w_c) t}].
\label{counter2}
\eeq

\medskip

\section{Matrix elements of $\bra FM|\mu_\pm|F'M'\ket$} 

For convenience to work with two-level H atom and Rb atom, 
here we derive the complete sets of equations.  

\subsection{H atom}

The $^{1}$H atom has a ground state split by a hyperfine interaction of
electron and nuclear spins ($L=0$, $S=1/2$, $I=1/2$)   
in two levels with different total angular moment $F=L+S+I$, which has two
values 0 and 1 as shown in Fig.~\ref{rb-level}b. 
The states are given by 
the elements of matrices $\bra FM|\mu_\pm|F'M'\ket$, which can be calculated
using
\begin{widetext}
\beq
|F=1,M=1\ket &=& |Sm_s=-1/2\ket|Im_i=-1/2\ket,\;\;
|F=1,M=-1\ket = |Sm_s=-1/2\ket|Im_i=-1/2\ket, \\
|F=1,M=0\ket &=& {1\over\sqrt{2}}|Sm_s=-1/2\ket|Im_i=1/2\ket + 
{1\over\sqrt{2}}|Sm_s=1/2\ket|Im_i=-1/2\ket, \\
|F=0,M=0\ket &=& 
{1\over\sqrt{2}}|Sm_s=-1/2\ket|Im_i=1/2\ket -
{1\over\sqrt{2}}|Sm_s=1/2\ket|Im_i=-1/2\ket.
\eeq
The nuclear magnetic moment can be neglected,
because it is much smaller than the electron magnetic
moment. We calculate thel elements of matrices as follows  
\be
\bra S,1/2|\mu_+|S,-1/2\ket =\bra S,1/2|\mu_-|S,-1/2\ket=\mu_S, 
\ee
where $\mu_S=g_S\mu_B$, $g_s = 2$, and $\mu_B$ is the Bohr's magneton. 
For example, 
\be
\bra F=1, M=1|\mu_+|F=1, M=0\ket = \bra S,1/2|\bra I,1/2|\mu_+
\left(
{1\over\sqrt{2}}|S,-1/2\ket |I,1/2\ket + 
{1\over\sqrt{2}}|S,1/2\ket |I,-1/2\ket
\right) = {\mu_S\over\sqrt{2}}. 
\ee
Other elements can be calculated similarly: 
\be
-\bra F=0, M=0|\mu_+|F=1, M=-1\ket = 
\bra F=0, M=0|\mu_+|F=1, M=1\ket = \mu_S,
\ee
\be
\bra F=1, M=0|\mu_+|F=1, M=-1\ket = 
\bra F=1, M=0|\mu_+|F=1, M=1\ket = \mu_S.
\ee
Then, Eqs.~(\ref{eq-rb}) can be explicitely written as
\beq
{d\over dt}\left( \begin{array}{c}
a_{1-1} \\
a_{10} \\
a_{11} \\
a_{00} 
\end{array}
\right) = \W_0\hat L
\left( \begin{array}{c}
a_{1-1} \\
a_{10} \\
a_{11} \\
a_{00} 
\end{array}
\right),
\eeq
where  $\W_0 = \mu_B B_s/\hbar$ and $\hat L$ is the following matrix
\[
\left( 
\begin{array}{cccc}
0             & \W e^{i\nu t}  & 0            & \W e^{-i(\nu+\w_c) t} \\
\W e^{-i\nu t} & 0             & \W e^{i\nu t} & 0 \\
0             & \W e^{-i\nu t} & 0            & \W e^{i(\nu-\w_c)} \\
\W e^{i(\nu+\w_c) t}& 0 & \W e^{-i(\nu-\w_c)}        & 0 
\end{array} 
\right).
\]  
\end{widetext}
This set of equations has been solved numerically. 

\subsection{Rb atom}

The $^{87}$Rb atom has a ground state split by a hyperfine interaction of
electron and nuclear spins ($L=0$, $S=1/2$, $I=3/2$)   
in two levels with different total angular moment $F=L+S+I$, which has two
values 1 and 2 as shown in Fig.~\ref{rb-level}c. 
The elements of matrices $\bra FM|\mu_\pm|F'M'\ket$ can be calculated
using
\begin{widetext}
\beq
|F=1,M=-1\ket &=& 
-{1\over2}|Sm_s=-1/2\ket|Im_i=-1/2\ket + 
{\sqrt{3}\over2}|Sm_s=1/2\ket|Im_i=-3/2\ket,\\
|F=1,M=0\ket &=& 
-{1\over\sqrt{2}}|Sm_s=-1/2\ket|Im_i=1/2\ket + 
{1\over\sqrt{2}}|Sm_s=1/2\ket|Im_i=-1/2\ket,\\
|F=1,M=1\ket &=& 
-{\sqrt{3}\over2}|Sm_s=-1/2\ket|Im_i=3/2\ket + 
{1\over2}|Sm_s=1/2\ket|Im_i=1/2\ket,\\
|F=2,M=-2\ket &=& |Sm_s=-1/2\ket|Im_i=-3/2\ket,\\
|F=2,M=-1\ket &=& 
{\sqrt{3}\over2}|Sm_s=-1/2\ket|Im_i=-1/2\ket + 
{1\over2}|Sm_s=1/2\ket|Im_i=-3/2\ket,\\
|F=2,M=0\ket &=& 
{1\over\sqrt{2}}|Sm_s=-1/2\ket|Im_i=1/2\ket + 
{1\over\sqrt{2}}|Sm_s=1/2\ket|Im_i=-1/2\ket,\\
|F=2,M=1\ket &=& 
{1\over2}|Sm_s=-1/2\ket|Im_i=3/2\ket + 
{\sqrt{3}\over2}|Sm_s=1/2\ket|Im_i=1/2\ket,\\
|F=2,M=2\ket &=& |Sm_s=1/2\ket|Im_i=3/2\ket.
\eeq
The nuclear magnetic moment  can be neglected,
because it is much smaller than the electron magnetic
moment. We calculate all elements of matrixes as follows
\be
\bra S,1/2|\mu_+|S,-1/2\ket =\bra S,1/2|\mu_-|S,-1/2\ket=\mu_S, 
\ee
where 
$\mu_S=g_S\mu_B$, $g_s = 2$, and $\mu_B$ is the Bohr's magneton. 
For example, 
\be
\bra F=2, M=2|\mu_+|F=1, M=1\ket = \bra S,1/2|\bra I,3/2|\mu_+
\left(-{\sqrt{3}\over2}|S,-1/2\ket |I,3/2\ket + 
{1\over2}|S,1/2\ket |I,1/2\ket\right) =-{\sqrt{3}\over2}\mu_S. 
\ee
Other elements can be calculated similarly 
\beq
\bra F=2, M=2|\mu_+|F=2, M=1\ket &=& {1\over2}\mu_S,
\bra F=2, M=2|\mu_+|F=1, M=1\ket = -{\sqrt{3}\over2}\mu_S,\\
\bra F=2, M=1|\mu_+|F=2, M=0\ket &=& {\sqrt{6}\over4}\mu_S,
\bra F=2, M=1|\mu_+|F=1, M=0\ket = -{\sqrt{6}\over4}\mu_S,\\
\bra F=2, M=0|\mu_+|F=2, M=-1\ket &=& {\sqrt{6}\over4}\mu_S,
\bra F=2, M=0|\mu_+|F=1, M=-1\ket = -{\sqrt{2}\over4}\mu_S,\\
\bra F=2, M=-1|\mu_+|F=2, M=-2\ket &=& {1\over2}\mu_S,
\bra F=1, M=1|\mu_+|F=1, M=0\ket = -{\sqrt{2}\over4}\mu_S,\\
\bra F=1, M=0|\mu_+|F=1, M=-1\ket &=& -{\sqrt{2}\over4}\mu_S,
\bra F=1, M=1|\mu_+|F=2, M=0\ket = {\sqrt{2}\over4}\mu_S,\\
\bra F=1, M=0|\mu_+|F=2, M=-1\ket &=& {\sqrt{6}\over4}\mu_S,
\bra F=1, M=-1|\mu_+|F=2, M=-2\ket = {\sqrt{3}\over2}\mu_S.
\eeq
Then, Eqs.~(\ref{eq-rb}) can be explicitely written as
\beq
{d\over dt}\left( \begin{array}{c}
a_{1-1} \\
a_{10} \\
a_{11} \\
a_{2-2} \\
a_{2-1} \\
a_{20} \\
a_{21} \\
a_{22}  
\end{array}
\right) = \W_0\hat L
\left( \begin{array}{c}
a_{1-1} \\
a_{10} \\
a_{11} \\
a_{2-2} \\
a_{2-1} \\
a_{20} \\
a_{21} \\
a_{22}  
\end{array}
\right),
\eeq
where $\W_0 = \mu_B B_s/\hbar$ and $\hat L$ is the following matrix
\[
\left( 
\begin{array}{cccccccc}
0 & -{\sqrt{2}\over4}e^{i\nu t} & 0 & {\sqrt{3}\over2}e^{-i(\nu+\w_c) t} & 0 & -{\sqrt{2}\over4}e^{i(\nu-\w_c) t} & 0 & 0 \\
-{\sqrt{2}\over4}e^{-i\nu t} & 0 & -{\sqrt{2}\over4}e^{i\nu t} & 0 & {\sqrt{6}\over4}e^{-i(\nu+\w_c) t} & 0 & -{\sqrt{6}\over4}e^{i(\nu-\w_c) t} & 0 \\
0 & -{\sqrt{2}\over4}e^{-i\nu t} & 0 & 0 & 0 & {\sqrt{2}\over4}e^{-i(\nu+\w_c) t} & 0 & -{\sqrt{3}\over2}e^{i(\nu-\w_c) t} \\
{\sqrt{3}\over2}e^{i(\nu+\w_c) t} & 0 & 0 & 0 & {1\over2}e^{i\nu t} & 0 & 0 & 0 \\
0 & {\sqrt{6}\over4}e^{i(\nu+\w_c) t} & 0 & {1\over2}e^{-i\nu t} & 0 & {\sqrt{6}\over4}e^{i\nu t} & 0 & 0 \\
-{\sqrt{2}\over4}e^{-i(\nu-\w_c) t} & 0 & {\sqrt{2}\over4}e^{i(\nu+\w_c) t} & 0 & {\sqrt{6}\over4}e^{-i\nu t} & 0 & {\sqrt{6}\over4}e^{i\nu t} & 0 \\
0 & -{\sqrt{6}\over4}e^{-i(\nu-\w_c) t} & 0 & 0 & 0 & {\sqrt{6}\over4}e^{-i\nu t} & 0 & {1\over2}e^{i\nu t} \\
0 & 0 & -{\sqrt{3}\over2}e^{-i(\nu-\w_c) t} & 0 & 0 & 0 & {1\over2}e^{-i\nu t} & 0 
\end{array} 
\right).
\] 
\end{widetext}

\end{document}